\documentclass{nature_frb}

\usepackage{url}
\usepackage{epsfig}
\usepackage{aas_macros}
\usepackage{amssymb,amsmath}
\usepackage{upgreek}
\usepackage{color}
\usepackage[normalem]{ulem}
\usepackage{verbatim}

\newcommand{\frb}{FRB\,121102}
\newcommand{\Sun}{\odot}
\newcommand{\Msun}{M$_\odot$}

\newcommand{\uJy}{$\mu$Jy}

\newcommand{\dmu}{\ensuremath{\mathrm{pc\; cm^{-3}}}}
\newcommand{\DM}{\rm DM}
\newcommand{\DMigm}{\rm DM_{\rm IGM}}

\newcommand{\arcsec}{\ensuremath{^{\prime\prime}}}
\newcommand{\arcmin}{\ensuremath{^{\prime}}}
\newcommand{\degree}{\ensuremath{^\circ}}

\newcommand{\hs}{\ensuremath{^{\mathrm{s}}}}
\newcommand{\hm}{\ensuremath{^{\mathrm{m}}}}
\newcommand{\hh}{\ensuremath{^{\mathrm{h}}}}

\long\def\symbolfootnote[#1]#2{\begingroup%
\def\thefootnote{\fnsymbol{footnote}}\footnote[#1]{#2}\endgroup}

\title{The direct localization of a fast radio burst and its host}

\author{
S.~Chatterjee$^{1}$,
C.~J.~Law$^{2}$,
R.~S.~Wharton$^{1}$,
S.~Burke-Spolaor$^{3,4,5}$,
J.~W.~T.~Hessels$^{6,7}$,
G.~C.~Bower$^{8}$,
J.~M.~Cordes$^{1}$,
S.~P.~Tendulkar$^{9}$,
C.~G.~Bassa$^{6}$,
P.~Demorest$^{3}$,
B.~J.~Butler$^{3}$,
A.~Seymour$^{10}$,
P.~Scholz$^{11}$,
M.~W.~Abruzzo$^{12}$,
S.~Bogdanov$^{13}$,
V.~M.~Kaspi$^{9}$,
A.~Keimpema$^{14}$,
T.~J.~W.~Lazio$^{15}$,
B.~Marcote$^{14}$
M.~A.~McLaughlin$^{4,5}$,
Z.~Paragi$^{14}$,
S.~M.~Ransom$^{16}$,
M.~Rupen$^{11}$,
L.~G.~Spitler$^{17}$,
\&
H.~J.~van Langevelde$^{14,18}$
}

\begin{document}

\maketitle

\noindent Published online by \emph{Nature} on 4 Jan 2017.  DOI: 10.1038/nature20797

\begin{affiliations}
\item Cornell Center for Astrophysics and Planetary Science and Department of Astronomy, Cornell University, Ithaca, NY 14853, USA
\item Department of Astronomy and Radio Astronomy Lab, University of California, Berkeley, CA 94720, USA
\item National Radio Astronomy Observatory, Socorro, NM 87801, USA
\item Department of Physics and Astronomy, West Virginia University, Morgantown, WV 26506, USA
\item Center for Gravitational Waves and Cosmology, West Virginia University, Chestnut Ridge Research Building, Morgantown, WV 26505
\item ASTRON, Netherlands Institute for Radio Astronomy, Postbus 2, 7990 AA, Dwingeloo, The Netherlands
\item Anton Pannekoek Institute for Astronomy, University of
   Amsterdam, Science Park 904, 1098 XH Amsterdam, The Netherlands
\item Academia Sinica Institute of Astronomy and Astrophysics, 645 N. A'ohoku Place, Hilo, HI 96720, USA
\item Department of Physics and McGill Space Institute, McGill University, 3600 University St., Montreal, QC H3A 2T8, Canada
\item Arecibo Observatory, HC3 Box 53995, Arecibo, PR 00612, USA
\item National Research Council of Canada, Herzberg Astronomy and Astrophysics, Dominion Radio Astrophysical Observatory, P.O. Box 248, Penticton, BC V2A 6J9, Canada
\item Haverford College, 370 Lancaster Ave, Haverford, PA 19041, USA
\item Columbia Astrophysics Laboratory, Columbia University,  New York, NY 10027, USA
\item Joint Institute for VLBI ERIC, Postbus 2, 7990 AA Dwingeloo, The Netherlands
\item Jet Propulsion Laboratory, California Institute of Technology, Pasadena, CA 91109, USA
\item National Radio Astronomy Observatory, Charlottesville, VA 22903, USA
\item Max-Planck-Institut f\"ur Radioastronomie, Auf dem H\"ugel 69, Bonn, D-53121, Germany
\item Sterrewacht Leiden, Leiden University, Postbus 9513, 2300 RA, Leiden, the Netherlands
\end{affiliations}

\newpage
\begin{abstract}
Fast radio bursts\cite{lbm+07,tsb+13} (FRBs) are astronomical radio flashes of unknown physical nature with durations of milliseconds.
Their dispersive arrival times suggest an extragalactic origin and imply radio luminosities orders of magnitude larger than any other kind of known short-duration radio transient\cite{cw16}.
Thus far, all FRBs have been detected with large single-dish telescopes with arcminute localizations, and attempts to identify their counterparts (source or host galaxy) have relied on contemporaneous variability of field sources\cite{kjb+16} or the presence of
peculiar field stars\cite{lsm14} or galaxies\cite{kjb+16}.
These attempts have not resulted in an unambiguous association\cite{wb16,vrm+16} with a host or multi-wavelength counterpart.
Here we report the sub-arcsecond localization of \frb, the only known repeating burst source\cite{sch+14,ssh+16a,ssh+16b,pjk+15}, using high-time-resolution radio interferometric observations that directly image the bursts themselves.
Our precise localization reveals that \frb\ originates within 100~milliarcseconds of a faint 180~\uJy\ persistent radio source with a continuum spectrum that is consistent with non-thermal emission, and a faint (twenty-fifth magnitude) optical counterpart.
The flux density of the persistent radio source varies by tens of percent on day timescales, and very long baseline radio interferometry yields an angular size less than 1.7 milliarcseconds.
Our observations are inconsistent with the fast radio burst having a Galactic origin or its source being located within a prominent star-forming galaxy. Instead, the source appears to be co-located with a low-luminosity active galactic nucleus or a previously unknown type of extragalactic source.
Localization and identification of a host or counterpart has been essential to understanding the origins and physics of other kinds of transient events,  including gamma-ray bursts\cite{mdk+97,bkd02} and tidal disruption events\cite{gmm+06}.
However, if other fast radio bursts have similarly faint radio and optical counterparts, our findings imply that direct sub-arcsecond localizations of FRBs may be the only way to provide reliable associations.

\end{abstract}

\newpage

The repetition of  bursts from \frb\cite{ssh+16a,ssh+16b} enabled a targeted interferometric localization campaign with the Karl G. Jansky Very Large Array (VLA) in concert with single-dish observations using
the 305-m William E. Gordon Telescope at the Arecibo Observatory. We searched for bursts in VLA  data with 5-ms sampling  using both beam-forming and imaging techniques\cite{lbb+15} (see Methods). In over 83~hr of VLA observations distributed over six months, we detected nine bursts from \frb\ in the 2.5--3.5~GHz band
with signal-to-noise ratios ranging from 10 to 150, all at a consistent sky position.
These bursts were initially detected with real-time de-dispersed imaging and confirmed by a beam-formed search (Figure 1).
From these detections, the average J2000 position of the burst source is right ascension $\alpha = 05\hh31\hm58.70\hs$, declination $\delta = +33\degree08\arcmin52.5\arcsec$, with an uncertainty of $\sim$0.1\arcsec, consistent with the Arecibo localization\cite{ssh+16a} but with three orders of magnitude better precision. The dispersion measure (DM) for each burst is consistent with the previously reported value\cite{ssh+16a} of 558.1$\pm$3.3~\dmu, with comparable DM uncertainties.
Three bursts detected at the VLA (2.5--3.5 GHz) had simultaneous coverage at Arecibo (1.1--1.7 GHz). After accounting for dispersion delay and light travel time, one burst is detected at both telescopes (Extended Data Table~1), but the other two show no emission in the Arecibo band, implying frequency structure at $\sim$1~GHz  scales. This provides new constraints on the broadband burst spectra, which previously have shown  highly variable  structure across the Arecibo band\cite{sch+14, ssh+16a, ssh+16b}.

Radio images at 3~GHz produced by integrating the VLA fast-sampled data reveal a continuum source within 0.1\arcsec\ of the burst position, which we refer to hereafter as the persistent source. A cumulative 3~GHz image (root mean square (r.m.s.) of $\sigma \approx$ 2~\uJy per beam; Figure~2) shows 68 other sources within a 5\arcmin\ radius, with a median flux density of 26~\uJy.
Given the match between burst positions and the continuum counterpart, we estimate a  probability $< 10^{-5}$ of chance coincidence.
The persistent source is detected in follow-up VLA observations over the entire  frequency range from 1 to 26~GHz. The radio spectrum is broadly consistent with non-thermal emission, though with significant deviation from a single power-law spectrum.
Imaging at 3~GHz over the campaign shows that the persistent source exhibits $\sim 10$\% variability on day timescales (Figure~2 and Extended Data Table~2). Variability in faint radio sources is common\cite{wb16,vrm+16}; of the 69 sources within a 5\arcmin\ radius, nine (including the persistent counterpart) were significantly variable (see Methods).
There is no apparent correlation between VLA detections of bursts from \frb\ and the flux density of the counterpart at that epoch (Figure~2, and Methods).

Observations with the European VLBI Network and the Very Long Baseline Array detect the persistent source and limit its size to $<1.7$~milliarcseconds (see Methods). The lower limit on the brightness temperature is $T_b > 8 \times 10^{6}$~K. The source has an integrated flux density consistent with that inferred at lower resolution in contemporaneous VLA imaging, indicating the absence of any significant flux on scales larger than a few milliarcseconds.

We have searched for counterparts at submillimeter, infrared, optical, and X-ray wavelengths using archival data and a series of new observations. A coincident unresolved optical source is detected in archival 2014 Keck data ($R_\mathrm{AB}=24.9\pm0.1$\,mag) and recently obtained Gemini data ($r_\mathrm{AB}=25.1\pm0.1$; Figure~2), with a chance coincidence probability $< 3.5\times10^{-4}$ (see Methods).
The source is undetected in archival infrared observations, in ALMA 230\,GHz observations, and in {\it XMM-Newton} and {\it Chandra} X-ray imaging (see Methods). The spectral energy distribution of the persistent source is compared in Figure~3 to some example spectra for known source types, none of which matches our observations well.

The observations reported here corroborate the strong arguments\cite{ssh+16b} against a Galactic location for the source.
As argued before, stellar radio flares can exhibit swept-frequency radio bursts on sub-second timescales\cite{mls+15a}, but they do not strictly adhere to the $\nu^{-2}$ dispersion law seen for \frb\cite{ssh+16a,ssh+16b}, nor are they expected to show constant apparent DM.
The source of the sizable DM excess, 3$\times$ the Galactic maximum
predicted by the NE2001 electron-density model\cite{cl02}, is not revealed as a HII region, a supernova remnant, or a pulsar-wind nebula in our Galaxy that would appear extended at radio, infrared, or H$\alpha$\cite{ssh+16b} wavelengths at our localized position.
Spitzer mid-infrared limits constrain sub-stellar objects with temperatures $>900$~K to be at distances of 70~pc or greater, and the Gemini detection sets a minimum distance of $\sim$1~kpc and 100~kpc for stars with effective temperatures greater than 3000~K and 5000~K, respectively, ruling out Galactic stars that could plausibly account for the DM and produce the radio continuum counterpart.
We conclude that \frb\ and its persistent counterpart do not correspond to any known class of Galactic source.

The simplest interpretation is that the burst source resides in a host galaxy which also contains the persistent radio counterpart.
If so, the DM of the burst source has contributions from the electron density in the Milky Way disk and halo\cite{cl02}, the intergalactic medium (IGM)\cite{ino04}, and the host galaxy.  We estimate $\DMigm = \DM - \DM_{\rm NE2001} - \DM_{\rm halo} - \DM_{\rm host} \approx 340$~pc~cm$^{-3}$ $- \DM_{\rm host}$  with $\DM_{\rm NE2001} = 188$~pc~cm$^{-3}$ and  $\DM_{\rm halo} \approx 30$~pc~cm$^{-3}$. The maximum redshift, for $\DM_{\rm host} = 0$, is $z_{\rm FRB} \lesssim 0.32$,  corresponding to a maximum luminosity distance of 1.7~Gpc.
Variance in the mapping of DM to redshift\cite{mcq14} ($\sigma_z = \sigma_{\rm DM} (dz/d\DM)\approx 0.1$) could increase the upper bound to $z \sim 0.42$.
Alternatively, a sizable host galaxy contribution could imply a low redshift and a negligible contribution from the IGM, although no such galaxy is apparent. Hereinafter we adopt $z_{\rm FRB} \lesssim 0.32$.

The faint optical detection and the non-detection at 230~GHz with ALMA imply a low star formation rate from any host galaxy.  For our ALMA 3$\sigma$ upper limit of 51~\uJy\ and a sub-mm spectral index of 4, we estimate the star formation rate\cite{cy99} to be less than 0.06 to 19 \Msun~yr$^{-1}$ for redshifts $z$ ranging from 0.01 to 0.32 (luminosity distances of 43 Mpc to 1.7 Gpc), respectively.  The implied absolute magnitude $\sim -16$ at $z = 0.32$ is similar to that of the Small Magellanic Cloud, whose mass $\sim 10^9$~\Msun  would comprise an upper limit on the host galaxy mass.

The compactness of the persistent radio source ($\lesssim$8~pc for $z \lesssim 0.32$) implies that it does not correspond to emission from an extended galaxy or a star forming region\cite{condon92}, although our brightness temperature limits do not require the emission to be coherent.
Its size and  spectrum appear consistent with a low-luminosity active galactic nucleus (AGN) but X-ray limits do not support this interpretation.
Young extragalactic supernova remnants\cite{ldt+06} can have brightness temperatures in excess of $10^7$~K but they typically have simple power-law spectra and exhibit stronger variability.

The burst source and persistent source have a projected separation $\lesssim$500~pc if  $z \lesssim 0.32$.  There are three broad interpretations of their relationship.
First, they may be unrelated objects harboured in a host galaxy,
such as a neutron star (or other compact object) and an AGN.
Alternatively, the two objects may interact, e.g., producing repeated bursts from a neutron star very close to an AGN\cite{pc15,cw16,lyu14}.
A third possibility is that they are a single source.
This could involve unprecedented bursts from an AGN\cite{rdv16} along with persistent synchrotron radiation;  or persistent emission might comprise high-rate bursts  too weak to detect individually, with bright detectable bursts forming a long tail of the  amplitude distribution. In this interpretation the difficulty in establishing any periodicity in the observed bursts\cite{ssh+16a,ssh+16b} may result from irregular beaming from a rotating compact object or extreme spin or orbital dynamics.
The Crab pulsar and some millisecond pulsars display bimodality\cite{lcu95, kt00} in giant and regular pulses. However, they show well-defined periodicities and have steep spectra
inconsistent with the spectrum of the persistent source that extends to at least 25~GHz.  Magnetars show broad spectra that extend beyond 100~GHz in a few cases but differ from the roll-off of the persistent source spectrum.

All things considered, we cannot favour any one of these interpretations.     Future comparison of spectra from  the persistent source and from  individual bursts could rule out the `single source' interpretation.
The proximity of the two sources and their physical relationship can  be probed by detecting a burst in VLBI observations or by using interstellar scintillations, which can resolve separations less than 1 milliarcsecond.

If other FRBs are similar to \frb, our discovery implies that direct subarcsecond localizations of bursts are so far the only secure way to find associations. The unremarkable nature of the counterparts to \frb\ suggests that efforts to identify other FRB counterparts in large error boxes will be difficult, and given the lack of correlation between the variability of the persistent source and the bursts, rapid post-FRB follow-up imaging in general may not be fruitful.

\clearpage

\begin{figure}
\begin{center}
\includegraphics[width=0.9\textwidth]{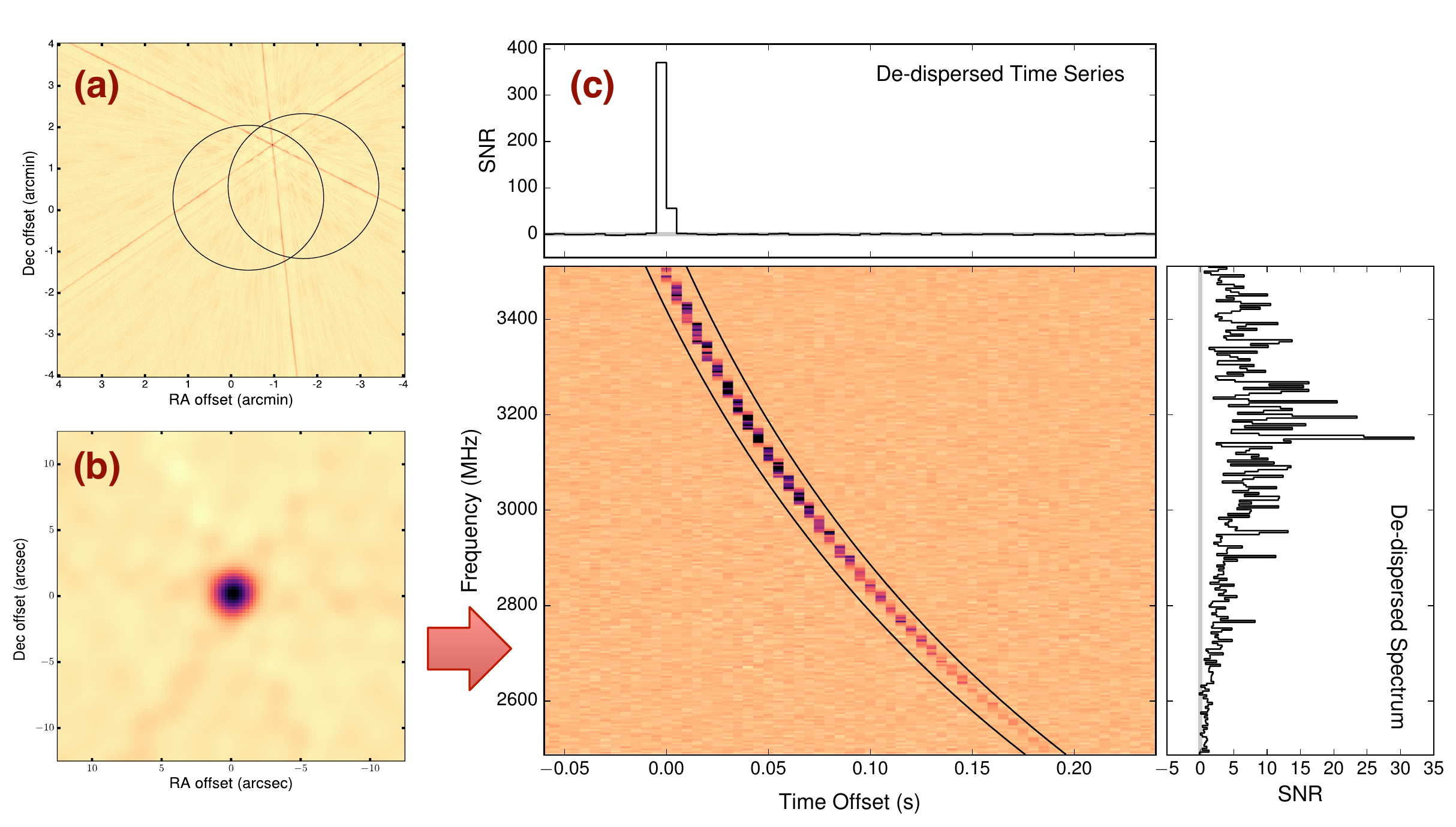}
\end{center}
\caption{VLA detection of \frb. (a) A 5-ms dispersion-corrected dirty image shows a burst from \frb\ at MJD 57633.67986367 (2016 Sep 02). The approximate localization uncertainty from previous Arecibo detections\cite{ssh+16a} (3\arcmin\ beam FWHM) is shown with overlapping circles. (b) A zoomed in portion of the above image, de-convolved and re-centered on the detection, showing the $\sim$0.1\arcsec\ localization of the burst. (c) Time-frequency data extracted from phased VLA visibilities at the burst location shows the $\nu^{-2}$ dispersive sweep of the burst.
The solid black lines illustrate the expected sweep for DM$=558$~\dmu. The de-dispersed lightcurve and spectra are projected to the upper and right panels, respectively.}
\vspace{-15cm}
\label{fig:detection}
\end{figure}

\clearpage

\begin{figure}
\begin{center}
\includegraphics[width=0.84\textwidth]{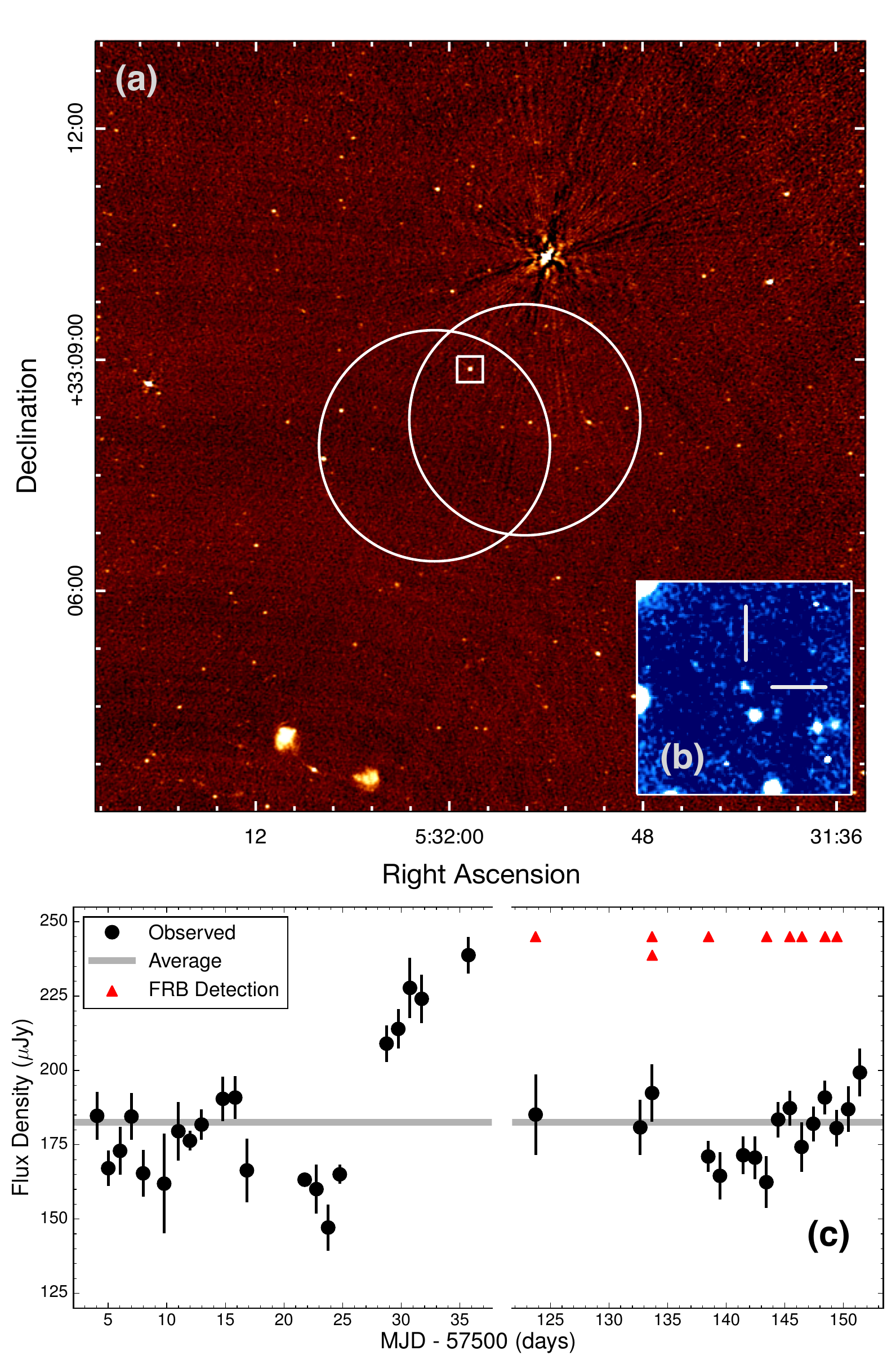}
\end{center}
\caption{Radio and optical images of the \frb\ field.
(a) VLA image at 3~GHz with a combination of array configurations. The image resolution is 2\arcsec\ and the RMS is $\sigma = 2$~\uJy~beam$^{-1}$. The Arecibo detection\cite{ssh+16a} uncertainty regions (3\arcmin\ beam FWHM) are indicated with overlapping white circles. The radio counterpart of the bursts detected at the VLA is highlighted with a 20\arcsec\ white square within the overlap region. (b) Gemini r-band image of the 20\arcsec\ square shows an optical counterpart
($r_\mathrm{AB}=25.1\pm0.1$~mag), as identified by the 5\arcsec\ bars.
(c) The lightcurve of the persistent radio source coincident with \frb\ over the course of the VLA campaign, indicating variability on timescales shorter than 1~day. Error bars are $1\sigma$. The average source flux density of $\sim$180~\uJy\ is marked, and the epochs at which bursts were detected at the VLA are indicated (red triangles). The variability of the persistent radio counterpart is uncorrelated with the detection of bursts (see Methods).}
\label{fig:field}
\end{figure}

\clearpage

\begin{figure}
\begin{center}
\includegraphics[width=0.9\textwidth]{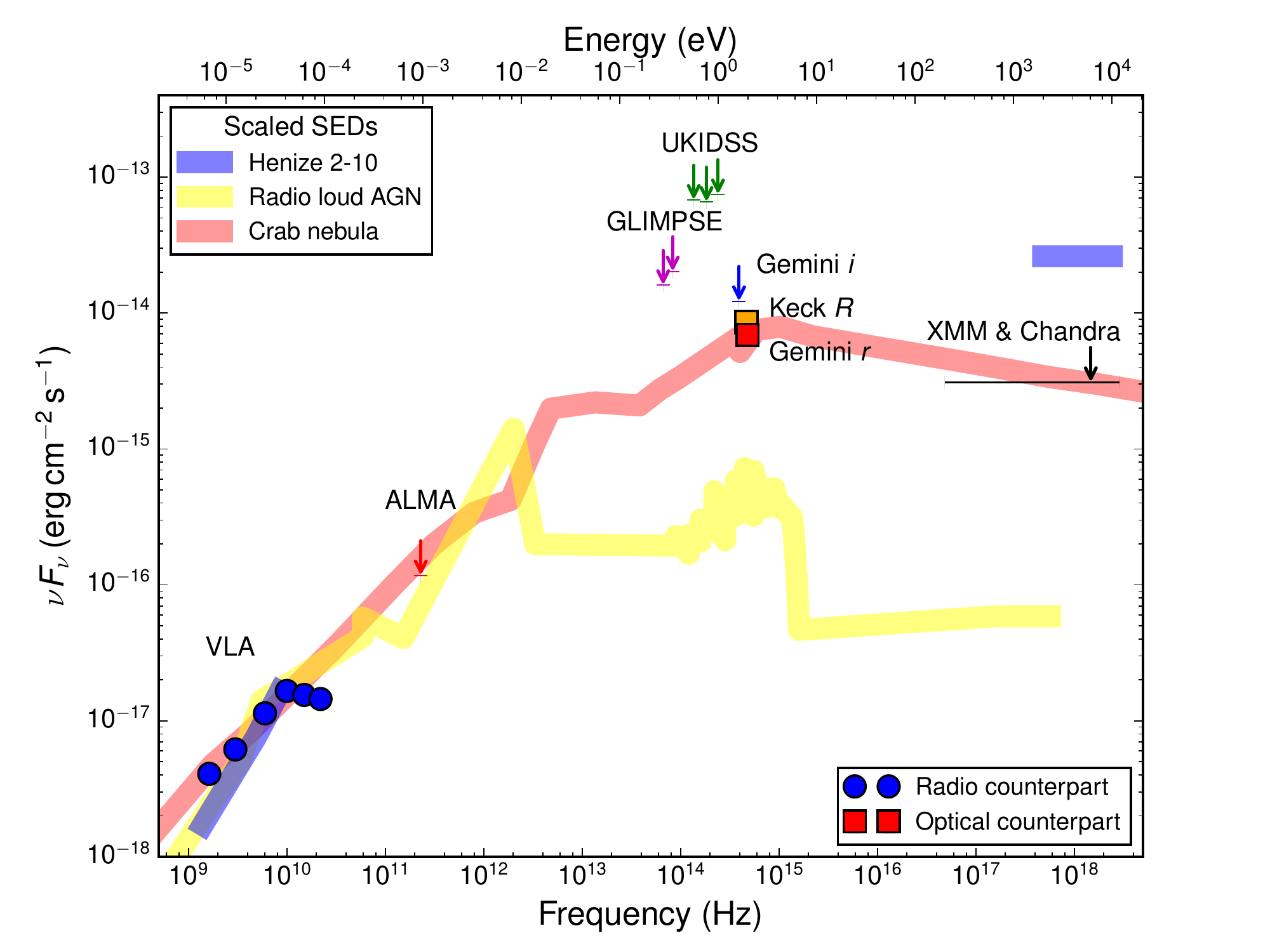}
\end{center}
\caption{Broadband spectral energy distribution (SED) of the counterpart.
Detections of the persistent radio source (blue circles), the optical counterpart (squares) and $5\sigma$ upper limits at various frequency bands (inverted arrows) are shown; see Methods for details.
SEDs of other radio point sources are scaled to match the radio flux density at 10~GHz and overlaid for comparison: (Blue) Low luminosity AGN in Henize 2-10, a star-forming dwarf Galaxy\cite{rd2012} placed at 25 Mpc; (Yellow) radio loud AGN QSO~2128$-$123\cite{ewm+1994} scaled by $10^{-4.3}$ to simulate a lower luminosity AGN and placed at 3 Gpc, and (Red) the Crab nebula\cite{bb2014} at 4~Mpc.
\label{fig:sed}}
\end{figure}

\clearpage


\clearpage
\section*{Methods}

\subsubsection*{Observation strategy.}
Detection and precise localization of an FRB requires $\sim$arcsecond angular resolution, $\sim$milli\-second time resolution, and $\sim$MHz frequency resolution.  In November 2015, we conducted 10~hours of fast dump (5~ms) observations of the \frb\ field with the VLA\cite{ssh+16b} at 1.6~GHz, with no burst detections. In April--May 2016, we observed for 40~hours at 3~GHz, and again detected no bursts with either our fast imaging or beam-forming pipeline (described below).  Detections of \frb\ with the 305-m Arecibo telescope\cite{ssh+16a,ssh+16b} suggested that VLA detections might be sensitivity limited, leading us to conduct a simultaneous observing campaign where Arecibo would identify a burst in the time domain and contemporaneous VLA observations would precisely localize it.  In practice, this proved to be unnecessary for VLA detection, but it provided a wider frequency band to characterize the burst spectra.

\subsubsection*{Arecibo observations.}
Arecibo observations used the L-wide receiver, which provides a frequency range of $1.15-1.73$~GHz. The PUPPI pulsar backend recorded full Stokes polarization information, with time and frequency resolutions of 10.24~$\mu$s and 1.5625~MHz, respectively. Each frequency channel was coherently dedispersed to 557~\dmu, thereby eliminating intra-channel dispersion smearing.

\subsubsection*{VLA fast-dump observations.}
The VLA fast-sampled interferometric data were recorded with 5~ms integration time, 256 channels, and bandwidth 1024~MHz centred at 3~GHz. We first detected \frb\ on 2016 August 23 with a signal-to-noise ratio S/N$\sim$35. Through 2016 September, we continued coordinated Arecibo and VLA observations, detecting another 8 bursts at the same location and DM. In total, we acquired $\sim$83~hrs of fast-dump interferometric observations in three sessions: 2015 November\cite{ssh+16b} at 1.6~GHz, 2016 April-May at 3~GHz, and 2016 August-September at 3~GHz with some observing at 6~GHz.

\subsubsection*{Millisecond imaging with fast-dump visibility data.}
During the coordinated campaign, all bursts were detected with real-time analysis within hours of the data being recorded by the \emph{realfast}\cite{lbb+15} system at the VLA.  The real-time processing system de-dispersed visibilities to a small range of values centred on DM = 557.0~pc~cm$^{-3}$. For each integration and DM value, the pipeline formed a Stokes I image on time scales from 5 to 80 ms and saved images with peak S/N greater than 7.4, a threshold based on the known false positive rate due to thermal noise.

All candidates were re-analyzed offline with improved calibration, data cleaning, and refined localization using both custom, Python-based software and CASA. We calibrated and imaged de-dispersed visibilities with a typical sensitivity ($1\sigma$) of 5~mJy in 5~ms. Extended Data Table~1 lists burst properties, and the brightest detection is shown in Figure~1. By fitting a model of the synthesized beam to an image of the burst, we measure burst locations with statistical errors better than 0.3\arcsec. However, the locations are affected by systematic errors at the level of about 1\% of the synthesized beam sizes, a modest effect that is evident when comparing the localizations of the first four VLA burst detections (beam sizes $\sim$2.5\arcsec $\times$ 2\arcsec) with the last five (beam sizes $\sim$1.3\arcsec $\times$ 0.8\arcsec).
Using just the last five burst centroids (with lower residual systematics due to the narrower beam), we find that the burst locations are consistent with the persistent radio continuum conterpart centroid (Extended Data Table~3 and Extended Data Figure~1). The radio continuum counterpart location is measured from the error-weighted mean of the location measured in deep imaging from 1 to 26~GHz (see below). The error in the offset is calculated from the quadrature sum of errors in each burst and the counterpart.

\subsubsection*{Beam-forming analysis with fast-dump visibility data.}
Beam-forming is complementary to millisecond imaging: instead of de-dispersing interferometric visibilities and searching for bursts in the image domain, the visibilities are summed with appropriate phasing to produce time-frequency data that can be searched for dispersed bursts. For the VLA observations, the calibration tables generated from time-averaged data (see below) were applied to the fast-dump visibility data, and custom Python software and existing CASA tools were used to extract time-frequency data per beam from a tiling of synthesized beams covering the search region.  The time-frequency data from each beam were then written to PSRFITS format and run through a single pulse search pipeline that used PRESTO pulsar processing tools.  Single pulse candidates from all the synthesized beams were jointly filtered to remove candidates that occurred simultaneously in many beams, as well as candidates that were narrow-band, as these were likely caused by radio frequency interference.  Diagnostic plots for the remaining candidates were examined by eye for bursts. The beam-forming pipeline was used to independently verify the times and positions of each of the VLA detected bursts.
For the example shown in Figure~1, the instrumental time resolution for the observations (5~ms) is much larger than both the intrinsic pulse width and the intra-channel DM smearing, leading to a pixelated appearance.

\subsubsection*{VLA imaging observations of the persistent counterpart.}
The 3~GHz VLA fast-dump observations were also averaged down to lower time resolution, calibrated using the standard VLA pipeline procedures with CASA\cite{casa}, and imaged at each epoch. Once the persistent counterpart to \frb\ had been identified, we used these per-epoch images to construct the light curve of the source, as well as a deep average image of the sky (Figure~2 and Extended Data Table~2).  The variability of the persistent radio counterpart is uncorrelated with the detection of bursts; the point biserial correlation coefficient between the detection (or not) of a burst and the flux density of the counterpart is $r=-0.054$, which would be exceeded by chance $\sim$75\% of the time. Of the 69 sources detected within a 5\arcmin\ radius, nine (including the persistent counterpart) showed significant variability, as measured by
$\chi^2_r = 1/(N-1) \; \Sigma_t (S_t - \bar{S})^2/\sigma_t^2 > 5.0$, where $S_t$ is the source flux density and $\sigma_t$ the image RMS at epoch $t$, and $\bar{S}$ is the epoch-averaged flux density.

We also acquired VLA imaging data covering a contiguous frequency range from 1 to 26 GHz. These observations utilized six separate receivers on the VLA: L- (1-2 GHz); S- (2-4 GHz); C- (4-8 GHz); X- (8-12 GHz); Ku- (12-18 GHz); and K-band (18-26 GHz). Observations were carried out on 2016 September 6 and 9, when the VLA was in the B-configuration, with maximum spacing between antennas of roughly 11 km (on September 9, a few antennas had been moved to their A-configuration locations).
A third epoch was observed on September 28, only at C-band, with the VLA in the most extended A-configuration.
Visibilities were dumped every 2 seconds, with channels of width either 1 or 2 MHz (depending on band). Calibration of the flux density scale was done using an
observation of 3C~48 at all bands\cite{pb13}, and the secondary calibrator J0555+3948 was used to monitor complex gain (amplitude and phase) fluctuations as a function of time throughout each of the observations. Standard calibration was done with the VLA calibration pipeline, and subsequent imaging done in both CASA and AIPS. Final flux densities were estimated by a number of techniques to provide a cross-check, including {\em imfit} in CASA, JMFIT in AIPS, summing up CLEAN component flux density, and summing up flux density in the image pixels. Positions were measured using JMFIT.
The two epochs (three for C-band) were imaged separately, and results between the two (three) were found to agree to within the uncertainties (Extended Data Figure 2), so visibility data from the two epochs (three for C-band) were combined together to make final images. Results are reported in Extended Data Table 3 and the measurements are plotted as part of the broad-band SED (Figure 3).

\subsubsection*{Very Long Baseline Interferometry with the European VLBI Network.}
The European VLBI Network (EVN) observed at 1.65~GHz in five epochs (2016 February 2, 10--11, 11--12 and 2016 May 24, May 25) for about two hours per session. The array included the 100~m Effelsberg, the 76~m Jodrell Bank, the 32~m Medicina, the 25~m Onsala, the 32~m Torun, the 25~m Westerbork (single dish), and the 305~m Arecibo telescopes. The data were streamed to the EVN Software Correlator (SFXC) at the Joint Institute for VLBI ERIC (JIVE) in Dwingeloo, at a data rate of 1024 Mbit/s (512 Mbit/s for Arecibo) in real time. The individual station voltages were recorded simultaneously as well. During the first epoch, the
ICRF source J0518+3306 was used as a phase-reference calibrator (separation from the field $\sim$2.9\degree) and observations were alternated between the field (8 minutes) and the calibrator (2 minutes). For subsequent epochs we used J0529+3209 as phase-reference calibrator (separation $\sim$1.1\degree) since it was proven to be sufficiently bright ($\sim$60~mJy) and compact for the EVN from the first epoch observations.

Following the VLA localization of \frb, we re-correlated all our observations with the phase center at the \frb\ position. The data were analyzed with AIPS following standard procedures, and the images were made with the Caltech Difmap package. We did not detect the persistent counterpart during the first epoch due to a combination of technical failures and the distant phase calibrator.
In the subsequent epochs we detected the persistent counterpart as a
slightly resolved source with typical peak brightness of about 100~\uJy~beam$^{-1}$ and integrated flux density of about 200~\uJy, and deconvolved source size of about 5 $\times$ 3~milliarcseconds at a position angle of $~$140\degree. The naturally-weighted beam size was about 18$\times$2.2 milliarcseconds in all cases, with a major axis position angle of $-$54\degree; the noise was 7\uJy~beam$^{-1}$.  Brightness temperature lower limits from the four successful epochs are $7 \times 10^6$~K.  (See Extended data Figure 2 and Table 3.)

\subsubsection*{Very Long Baseline Interferometry with the Very Long Baseline Array.}

The NRAO Very Long Baseline Array (VLBA) observed on 2016 Sep 09, 16 with 8 hour tracks per epoch. First epoch observations were at $1.392-1.680$~GHz, with a synthesized beam size of 11.3 $\times$ 5.0~milliarcseconds at a position angle = 163.7\degree.
Second epoch observations were at $4.852-5.076$~GHz (beam size 2.74 $\times$ 1.43~milliarcseconds at a position angle = 174.8\degree).
A total recording bandwidth of 2~Gbps with dual circular polarizations was obtained for each observation.  As in the EVN observations, the compact calibrator J0529+3209 was used to provide phase referencing solutions for \frb.  Standard interferometric calibrations were applied using AIPS.  Images of the field achieved 17 and 12~\uJy~beam$^{-1}$ RMS at 1.5 and 5.0~GHz, respectively. The persistent counterpart to \frb\ was clearly detected in both observations with partially resolved compact structure.
At 1.5~GHz, two-dimensional Gaussian deconvolution yields a size of 4.6$\times$3.3~milliarcseconds, while the 5.0~GHz upper limit on the deconvolved size is $<$1.73~milliarcseconds.
Brightness temperature lower limits from the two epochs are $8 \times 10^6$ and $3 \times 10^6$~K, respectively.  (See Extended data Figure 2 and Table 3.)

\subsubsection*{Atacama Large Millimeter Array observations.}
The Atacama Large Millimeter and Submillimeter Array (ALMA) observed on 2016 Sep 15, using Band 6 and covering 8~GHz of bandwidth in the range $220-240$~GHz (with 2-MHz channels).
We used 38 antennas in the C40-6 configuration, yielding a resolution of 0.32\arcsec $\times$ 0.13\arcsec.
Calibration and imaging was provided by the ALMA observatory, and done using CASA via the ALMA pipeline.
The image RMS noise level was 17~\uJy~beam$^{-1}$ and did not reveal any significant sources.

\subsubsection*{Optical and infrared imaging}
We used the following optical imaging data: 630-s Keck $R$-band image from 2014 November 19, 120-s Gemini $i$-band image (2016 March 17), and 1250-s Gemini $r$-band image (2016 October 24 and 25). The data were reduced with a combination of \texttt{IRAF}, \texttt{IDL} and {Python} tools. We used the URAT1 catalog\cite{zfs+2015} as an astrometric reference frame. The astrometric errors were $<80-90$~mas (RMS) in all frames. We used the IPHAS photometry\cite{dgi+2005} to measure the zero-point correction for the coadded images. A counterpart to \frb\ is detected in archival Keck image at $R_\mathrm{AB}=24.9\pm0.1$~mag and in Gemini GMOS $r$-band image at $r_\mathrm{AB}=25.1\pm0.1$~mag, consistent with being point-like in 0.7\arcsec\ seeing. A non-detection in the $i$-band image yields an upper limit of $i>24$. The $r$-band centroid position was measured to be $\alpha = 05\hh31\hm58.69\hs$, $\delta = +33\degree08\arcmin52.51\arcsec$ (J2000) with an astrometric error of $\approx 100$\,mas, consistent with the $R$-band position. We measure a stellar density of $1.12\times10^{-2}\,\mathrm{arcsec^{-2}}$ for $r_\mathrm{AB}<25.1$. Thus, the chance coincidence probability of finding the optical counterpart within a 100~mas radius of the radio position is $< 3.5 \times 10^{-4}$. In regions around the \frb, we derive an upper limit of $r < 26.2$~AB mag~arcsec$^{-2}$ (5~$\sigma$) for any diffuse emission from an extended galaxy, ruling out most of the massive ultra-low surface brightness galaxies\cite{vda+2015}. These galaxies are as large as the Milky Way and would have been more than 5\arcsec\ in diameter if placed at a $z_{DM}=0.3$. Smoothing the image with a 5\arcsec\ FWHM Gaussian kernel reveals no significant emission on those scales.

The counterpart is not detected in near and mid-infrared observations from the UKIDSS\cite{lwa+07} and Deep GLIMPSE\cite{bcb+03,cbm+09} surveys with upper limits of $J=19.8$, $H=19.0$ and $K=18.0$ for UKIDSS and 17.8 and 17.3 for the GLIMPSE 3.6 and 4.5~$\upmu$m bands. At the location of \frb, the total $V$-band absorption, as determined from the COBE/DIRBE dust maps\cite{sfd98}, is 2.42\,mag. We use published extinction coefficients\cite{sf11} to correct for absorption in the other bands. Published zeropoints and effective wavelengths\cite{bcp98,fig+96,hwlh06} and the IRAC Instrument Handbook v2.1.2 were used to obtain the flux density measurements and limits shown in the broadband spectrum of the persistent counterpart (Figure~3).

\subsubsection*{X-ray Imaging with {\it XMM-Newton} and {\it Chandra} X-ray Observatory}
X-ray observations were done with {\it XMM-Newton} (IDs 0790180201, 0790180501, 0792382801, and 0792382901) and the {\it Chandra} X-ray Observatory (ID 18717). The cameras aboard {\it XMM-Newton} consist of one EPIC-pn\cite{sbd+01} and two EPIC-MOS\cite{taa+01} CCD arrays. The {\it Chandra} observation used the ACIS-S3 detector in TE mode. Two {\it XMM-Newton} observations occurred in 2016 February--March, before we achieved our precise localization, with pn in Large Window mode and the MOS cameras in Full Frame mode. Two more observations were performed in 2016 September with the pn camera in Small Window mode and the MOS cameras in Timing mode. A 40~ks {\it Chandra} observation was performed in 2015 November. In the first two {\it XMM-Newton} observations, the pn data were not usable for imaging \frb\ as it was positioned at the edge of a CCD chip. The MOS Timing mode observations are also not usable for imaging purposes. We therefore used 41~ks of pn data from 2016 September and 60~ks of MOS data from 2016 February--March for X-ray imaging.

We used standard tools from the XMM Science Analysis System (SAS) version 14.0, HEASoft version 6.19 and CAIO version 4.7 to reduce the data.
The {\it XMM-Newton} images were mosaicked together using {\tt emosaic} using exposure maps from {\tt eexpmap}.
The number of counts in a $18^{\prime\prime}$ and $1^{\prime\prime}$ radius circular region, for {\it XMM-Newton} and {\it Chandra}, respectively, centered at the position of \frb\ were compared to several randomly selected background regions. No significant deviation from the background was found; the 5$\sigma$ count rate limits are $<3\times10^{-4}$~counts~s$^{-1}$ and $<2\times10^{-4}$~counts~s$^{-1}$.
To place a flux limit we assume a photoelectrically absorbed power-law spectrum with a spectral index of $\Gamma=2$ and $N_\mathrm{H}= 1.7\times10^{22}$\,cm$^{-2}$  (the hydrogen column density implied by the DM--$N_\mathrm{H}$ relation\cite{hnk13}). Taking into account the telescopes' energy-dependent effective area and Poisson statistics, we place a 5$\sigma$ limit of $5\times10^{-15}$\,erg\,s$^{-1}$\,cm$^{-2}$ at 0.5--10~keV on an X-ray point source at the location of \frb\ using the mosaicked  {\it XMM-Newton} image. Using the same procedure on the {\it Chandra} image also results in a limit of $5\times10^{-15}$\,erg\,s$^{-1}$\,cm$^{-2}$ at 0.5--10~keV.

\subsubsection*{Observational constraints on \frb\ and its persistent counterpart.}

Our observations support the conclusion that no Galactic source can explain the observed DM excess.  If the compact counterpart contributes the excess DM over the maximum predicted by NE2001 along this line of sight, the requirement that it be optically thin\cite{ssh+16b} at 1.4~GHz implies a lower limit on its size (L $> 0.03$~pc), and hence the source distance. The VLBA and EVN compactness limits ($<1.7$~mas in any case, ignoring scattering contributions to angular extent) imply a minimum distance $>3.6$~Mpc, far beyond our Galaxy.
The absence of an X-ray detection constrains an AGN counterpart. The fundamental plane relation\cite{kfc06} between radio and X-ray luminosities and the black hole mass predicts that X-ray emission should be detected for black hole systems with $z < 0.32$ and $M_{BH} < 10^9\, M_\Sun$.  However, not all AGN follow this relationship, including radio-loud AGN and systems with jet-ISM interactions.  Radio-loud AGN are likely excluded based on the low radio luminosity $L_R\approx 3 \times 10^{41}$ erg s$^{-1}$ at $z=0.32$. A $10^6\, M_\Sun$ black hole, which is plausible given the $\sim 10^9 M_\Sun$ stellar mass upper limit, would have to accrete at $<10^{-2}$ below the Eddington rate to match the X-ray upper limit.
Our observations are also inconsistent with a young radio supernova remnant, which is typically variable on a time scale of months and associated with star formation\cite{ldt+06}.

For a nominal Gpc distance $D$ corresponding to redshifts $z\lesssim 0.3$, the received fluence $A_{\nu}$ from each burst implies  a burst energy
$$E_{\rm burst} = 4\pi D^2 (\delta\Omega/4\pi) A_{\nu} \Delta\nu
\approx 10^{38}\, {\rm erg}\,(\delta\Omega/4\pi) D_{\rm Gpc}^2  (A_{\nu} / 0.1\ {\rm Jy\ ms}) \Delta\nu_{\rm GHz}.$$
The unknown emission solid angle $\delta\Omega$
could be very small due to relativistic beaming, and together with a distance possibly much smaller than 1~Gpc, could reduce the energy requirement significantly.  However, the {\it total} energy emitted could be larger depending on the duration of the emission in the source frame and other model-dependent details.
Either way, the burst energies from \frb\ are not inconsistent with those that might be expected from the magnetosphere of a compact object\cite{cw16}.

\subsubsection*{Data availability.}
All relevant data are available from the authors.
VLA visibility data selected for times centered on each of the nine bursts (including Figure~1) are available at: \url{https://doi.org/10.7910/DVN/TLDKXG}. Data presented in Figures 2(c), 3, and Extended data Figures 1 and 2 are included with the manuscript.
The observational data presented here are available from public archives under the following project codes.
VLA fast dump observations: 15B-378, 16A-459, 16A-496;
VLA imaging: 16B-385;
ALMA: ADS/JAO.ALMA\#2015.A.00025.S;
VLBA: 16B-389, 16B-406;
EVN: RP024;
Gemini: GN-2016A-FT-5, GN-2016B-DD-2.

\subsubsection*{Code availability.}
Computational notebooks for reproducing the burst position analysis are at \url{http://github.com/caseyjlaw/FRB121102}.
The code used to analyse the data and observations reported here is available at the following sites:\\
{\em Realfast} (\url{http://realfast.io}),\\
{RTPipe} (\url{https://github.com/caseyjlaw/rtpipe}),\\
{SDMPy} (\url{http://github.com/demorest/sdmpy}).\\
Other standard data reduction packages (AIPS, CASA, Difmap, PyRAF, XMM SAS, HEASoft, CIAO, PRESTO) are available at their respective websites.

\clearpage


\begin{addendum}

\item The National Radio Astronomy Observatory is a facility of the National Science Foundation operated under cooperative agreement by Associated Universities, Inc.
We thank the staff at the NRAO for their continued support of these observations, especially with scheduling and computational infrastructure.
The Arecibo Observatory is operated by SRI International under a cooperative agreement with the National Science Foundation (AST-1100968), and in alliance with Ana G.~M\'{e}ndez-Universidad Metropolitana, and the Universities Space Research Association.
We thank the staff at Arecibo for their support and dedication that enabled these observations. Further acknowledgements of telescope facilities and funding agencies are included in the Supplementary Material.

\item[Contributions]
S.C. was PI of the localization campaign described here.
C.J.L. and S.B-S. are PIs of the {\em realfast} project and performed the analysis that achieved the first VLA burst detections.
S.C., C.J.L., R.S.W., S.B-S., G.C.B., B.B., and P.D. performed
detailed analysis of the VLA data.
S.B-S. and B.B. led the analysis
of the VLA multi-band spectral data.  J.W.T.H. was PI of the EVN
observations, which were analyzed by Z.P. and B.M. G.C.B. was PI of
the VLBA observations, and led their analysis.  J.W.T.H., A.S. and
L.G.S. led the execution and analysis of the parallel Arecibo observing
campaign. P.D. led the commissioning of fast-sampled VLA observing modes.
S.C. was PI of the ALMA observations.
P.S. was PI of the X-ray observations, and
performed the X-ray analysis, along with S.B.
S.P.T. was PI of the Gemini observations, and
along with C.G.B. led the analysis of Keck, Gemini, and
archival UKIDSS and GLIMPSE data. S.C. and
C.J.L. led the writing of the manuscript, with significant
contributions from J.M.C. and J.W.T.H.  All authors
contributed substantially to the interpretation of the analysis
results and to the final version of the manuscript.

\item[Competing Interests] The authors declare that they have no
  competing financial interests.

\item[Correspondence] Correspondence and requests for materials should
  be addressed to S.C.\\
  (email: shami.chatterjee@cornell.edu).

\end{addendum}

\clearpage
\section*{Extended Data}

\bigskip\noindent
Extended Data Table 1: {\bf VLA detections of bursts from \frb\ and Arecibo constraints.}
Dates are all in the year 2016. Position offsets $\Delta$RA and $\Delta$Dec are measured from RA = 05\hh31\hm58\hs, Dec = +33\degree08\arcmin52\arcsec (J2000). Topocentric burst MJDs are reported as offsets from MJD~57620 and are dedispersed to the top of the VLA band at 3.5~GHz. The flux densities and signal-to-noise ratios are estimated from a 5~ms visibility integration, leading to an underestimate since the burst durations are typically shorter in Arecibo detections. Instantaneous beam sizes are listed. Bursts with simultaneous coverage at Arecibo at 1.4~GHz are indicated with estimated detection peak flux density in a 5~ms integration, or 5$\sigma$ upper limits for non-detections.

\begin{table}
{\label{tab:bursts}}
\medskip
\begin{tabular}{llllllllll}
\hline
\hline
Date & $\Delta$MJD & $\Delta$RA & RA & $\Delta$Dec & Dec & Beam & S/N & VLA flux & AO flux \\
& from & & error & & error & size & ratio & density & density \\
 & 57620.0 & (s) & (\arcsec) & (\arcsec) & (\arcsec) & (\arcsec, \arcsec) & & (mJy) & (mJy) \\[1mm]
\hline
23 Aug & 3.74402686 & 0.68 & 0.05  & 0.65 & 0.07 & 2.8 $\times$ 2.3 & 27$^\dag$ & 120 & --\\
02 Sep & 13.67986367 & 0.70 & 0.01 & 0.48 & 0.01 & 2.5 $\times$ 2.3 & 149 & 670 & -- \\
02 Sep & 13.69515938 & 0.70 & 0.3  & 0.3 & 0.3 & 2.6 $\times$ 2.3 & 7$^\dag$ & 25 & --  \\
07 Sep & 18.49937435 & 0.71 & 0.07 & 0.50 & 0.07 & 1.9 $\times$ 1.7 & 13 & 63 & -- \\
12 Sep & 23.45730263 & 0.70 & 0.01 & 0.55 & 0.03 & 1.9 $\times$ 0.9 & 66 & 326 & -- \\
14 Sep & 25.42958602 & 0.69 & 0.07 & 0.54 & 0.12 & 1.3 $\times$ 0.7 & 10 & 39 & $\lesssim$6 \\
15 Sep & 26.46600650 & 0.70 & 0.05 & 0.56 & 0.05 & 1.2 $\times$ 0.7 & 10$^\dag$  & 50 & -- \\
17 Sep & 28.43691490 & 0.70 & 0.03 & 0.57 & 0.03 & 1.3 $\times$ 0.8 & 18 & 86 & $\sim$14 \\
18 Sep & 29.45175697 & 0.70 & 0.02 & 0.49 & 0.02 & 1.3 $\times$ 0.8 & 34 & 159 & $\lesssim$6 \\
\hline
\hline
\end{tabular}
\end{table}

$^\dag$
The real-time analysis of events on 23 August, 02 September, and 15 September resulted in detections with S/N ratio of 35, 16, and 16, respectively.
These differences are due to different calibration and flagging approaches between the real-time and offline analyses.  The real-time analysis does not include flux density scale calibration, so the offline analysis gives the best estimate of their flux density.

\clearpage
\bigskip\noindent
Extended Data Table 2: {\bf VLA 3~GHz observations of the persistent counterpart to \frb\ over time.}
Most observations were acquired during array reconfigurations (C$\rightarrow$CnB; CnB$\rightarrow$B; B$\rightarrow$A).
Horizontal lines denote changes in array configuration, as indicated by the changes in the synthesized beam size.

\begin{table}
{\label{tab:frblc}}
\medskip
\begin{tabular}{lllll}
\hline
\hline
Date & MJD & Beam size & Flux density & Bursts \\
(in 2016)  & & (\arcsec, \arcsec) & ($\mu$Jy) & detected \\
\hline
26 Apr & 57504.0565  & 6.25 $\times$ 2.42 &184.72  $\pm$  7.97  & 0   \\
27 Apr & 57505.0100  & 6.33 $\times$ 1.98 &167.08  $\pm$  5.91  & 0   \\
28 Apr & 57506.0181  & 6.44 $\times$ 1.96 &172.93  $\pm$  8.01  & 0   \\
28 Apr & 57506.9871  & 6.14 $\times$ 1.98 &184.49  $\pm$  7.90  & 0   \\
29 Apr & 57507.9843  & 6.12 $\times$ 1.98 &165.37  $\pm$  7.81  & 0   \\ \hline
01 May & 57509.7766  & 8.21 $\times$ 1.89 &161.90  $\pm$  16.79  & 0  \\
02 May & 57510.9890  & 6.44 $\times$ 2.02 &179.54  $\pm$  9.91  & 0   \\
03 May & 57511.9746  & 6.29 $\times$ 2.02 &176.34  $\pm$  3.36  & 0   \\
04 May & 57512.9725  & 6.29 $\times$ 1.95 &181.78  $\pm$  5.10  & 0   \\
06 May & 57514.7883  & 7.12 $\times$ 1.96 &190.43  $\pm$  7.43  & 0   \\
07 May & 57515.8362  & 6.23 $\times$ 2.00 &190.86  $\pm$  7.13  & 0   \\
08 May & 57516.8335  & 6.19 $\times$ 2.03 &166.36  $\pm$  10.72  & 0  \\
13 May & 57521.7640  & 6.83 $\times$ 1.95 &163.23  $\pm$  2.23  & 0   \\
14 May & 57522.7649  & 6.78 $\times$ 1.95 &160.06  $\pm$  8.22  & 0   \\
15 May & 57523.7658  & 6.61 $\times$ 1.95 &147.13  $\pm$  7.77  & 0   \\
16 May & 57524.7550  & 8.31 $\times$ 1.94 &165.03  $\pm$  3.23  & 0   \\
20 May & 57528.7452  & 1.79 $\times$ 1.58 &209.01  $\pm$  6.14  & 0   \\
21 May & 57529.7440  & 1.80 $\times$ 1.60 &213.95  $\pm$  6.65  & 0   \\
22 May & 57530.7439  & 1.76 $\times$ 1.60 &227.76  $\pm$  10.13  & 0  \\
23 May & 57531.7441  & 1.78 $\times$ 1.60 &224.08  $\pm$  8.14  & 0   \\
27 May & 57535.7339  & 1.77 $\times$ 1.60 &238.78  $\pm$  6.18  & 0   \\ \hline
23 Aug & 57623.7454  & 2.12 $\times$ 1.68 &185.15  $\pm$  13.56  & 1  \\
01 Sep & 57632.6730  & 2.22 $\times$ 1.71 &180.82  $\pm$  9.24  & 0   \\
02 Sep & 57633.6800  & 1.83 $\times$ 1.66 &192.39  $\pm$  9.69  & 2   \\
07 Sep & 57638.4685  & 1.90 $\times$ 0.61 &171.03  $\pm$  5.21  & 1   \\
08 Sep & 57639.4684  & 2.02 $\times$ 0.60 &164.53  $\pm$  7.96  & 0   \\
10 Sep & 57641.4579  & 1.52 $\times$ 0.59 &171.46  $\pm$  6.28  & 0   \\
11 Sep & 57642.4581  & 1.55 $\times$ 0.59 &170.65  $\pm$  7.21  & 0   \\
12 Sep & 57643.4272  & 1.56 $\times$ 0.58 &162.40  $\pm$  8.72  & 1   \\
13 Sep & 57644.4332  & 1.04 $\times$ 0.54 &183.47  $\pm$  5.96  & 0   \\
14 Sep & 57645.4307  & 1.01 $\times$ 0.49 &187.32  $\pm$  5.92  & 1   \\
15 Sep & 57646.4280  & 0.99 $\times$ 0.49 &174.24  $\pm$  8.38  & 1   \\
16 Sep & 57647.4245  & 1.01 $\times$ 0.48 &182.05  $\pm$  5.87  & 0   \\
17 Sep & 57648.4161  & 1.03 $\times$ 0.49 &190.88  $\pm$  5.69  & 1   \\
18 Sep & 57649.4162  & 1.02 $\times$ 0.49 &180.58  $\pm$  6.16  & 1   \\
19 Sep & 57650.4058  & 1.07 $\times$ 0.51 &186.93  $\pm$  7.69  & 0   \\
20 Sep & 57651.4058  & 1.02 $\times$ 0.49 &199.29  $\pm$  8.12  & 0   \\
\hline
\hline
\end{tabular}
\end{table}

\clearpage
\bigskip\noindent
Extended Data Table 3: {\bf Flux density and position measurements of the persistent counterpart to \frb.} We report the VLA radio spectrum with continuous frequency coverage from 1 to 25~GHz, along with detection positions and a weighted average position that is consistent with the detected burst positions to within 0.1\arcsec. We also list VLBA, EVN, and Gemini detection positions.
Position offsets $\Delta$RA and $\Delta$Dec are measured from a nominal RA = 05\hh31\hm58\hs, Dec = +33\degree08\arcmin52\arcsec (J2000).
The Gemini $r$-band detection position is also included.
1-$\sigma$ errors are quoted in all cases.

\begin{table}
{\label{tab:vla-sed}}
\medskip
\begin{tabular}{lllll}
\hline
\hline
Telescope  & Frequency & Flux density & $\Delta$RA & $\Delta$Dec \\
           & (GHz)     & ($\mu$Jy)    & (s)        &  ($\arcsec$) \\[1mm]
\hline
\rule{0pt}{2ex} VLA       &  1.63 & 250 $\pm$ 39 & 0.694 $\pm$ 0.018 & 0.43 $\pm$ 0.26 \\
          &  3.0  & 206 $\pm$ 17 & 0.705 $\pm$ 0.005 & 0.43 $\pm$ 0.07 \\
          &  6.0  & 203 $\pm$  7 & 0.701 $\pm$ 0.005 & 0.54 $\pm$ 0.01 \\
          & 10.0  & 166 $\pm$  9 & 0.701 $\pm$ 0.001 & 0.54 $\pm$ 0.02 \\
          & 15.0  & 103 $\pm$  7 & 0.691 $\pm$ 0.002 & 0.65 $\pm$ 0.02 \\
          & 22.0  &  66 $\pm$  7 & 0.699 $\pm$ 0.001 & 0.56 $\pm$ 0.01 \\
\multicolumn{3}{l}{\rule{0pt}{2ex} VLA Weighted Average} & 0.6998 $\pm$ 0.0004 & 0.548 $\pm$ 0.006 \\[1mm]
\hline
\rule{0pt}{2ex} EVN       & 1.67  & 200 $\pm$ 20   & 0.70150 $\pm$  0.00003 & 0.5505 $\pm$ 0.0003 \\[1mm]
\hline
\rule{0pt}{2ex} VLBA    & 1.55  & 218 $\pm$ 38   & 0.70159  $\pm$  0.00002  & 0.5508 $\pm$ 0.0006 \\
        & 4.98  & 151 $\pm$ 19   & 0.701530 $\pm$  0.000003 & 0.54952 $\pm$ 0.00009\\[1mm]
\hline
\rule{0pt}{2ex} Gemini       &       &                & 0.69      $\pm$ 0.01      &  0.51 $\pm$ 0.10 \\
\hline
\hline
\end{tabular}
\end{table}

\clearpage
\bigskip\noindent
Extended Data Figure 1: {\bf The offset of \frb\ from the persistent counterpart.} Five bursts detected at the VLA with the highest resolution (A-array, 3~GHz) are plotted, with epoch indicated by MJD values. The (RA, Dec) coordinate difference (burst relative to counterpart) is shown with an ellipse indicating the $1\sigma$ error calculated as the quadrature sum of errors in the two sources. VLBA and EVN positions are indicated, with $1\sigma$ errors smaller than the symbols. The centroid of the Gemini optical counterpart is shown (red dot) with an estimated $1\sigma$ error circle of 100~mas (red) from fitting and radio-optical frame tie uncertainties.

\begin{figure}
\begin{center}
\includegraphics[width=0.95\textwidth]{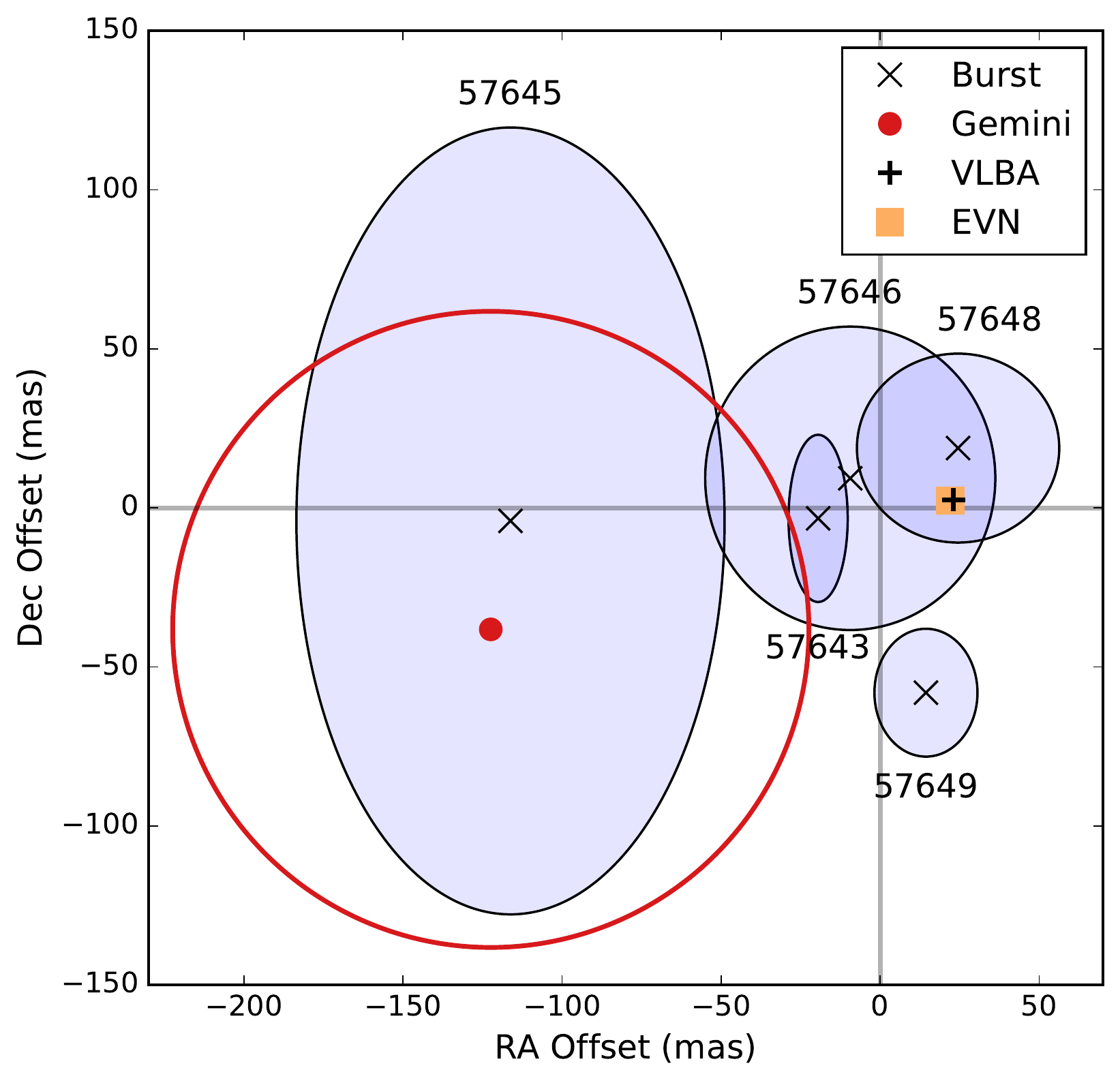}
\end{center}
\label{fig:centroid}
\end{figure}

\clearpage
\bigskip\noindent
Extended Data Figure 2: {\bf VLA spectrum of the persistent counterpart to \frb.}
The integrated flux density is plotted for each epoch of observation (listed by MJD) over a frequency range $\nu$ from 1 to 25~GHz. The spectrum is non-thermal and inconsistent with a single power law.

\begin{figure}
\begin{center}
\includegraphics[width=0.95\textwidth]{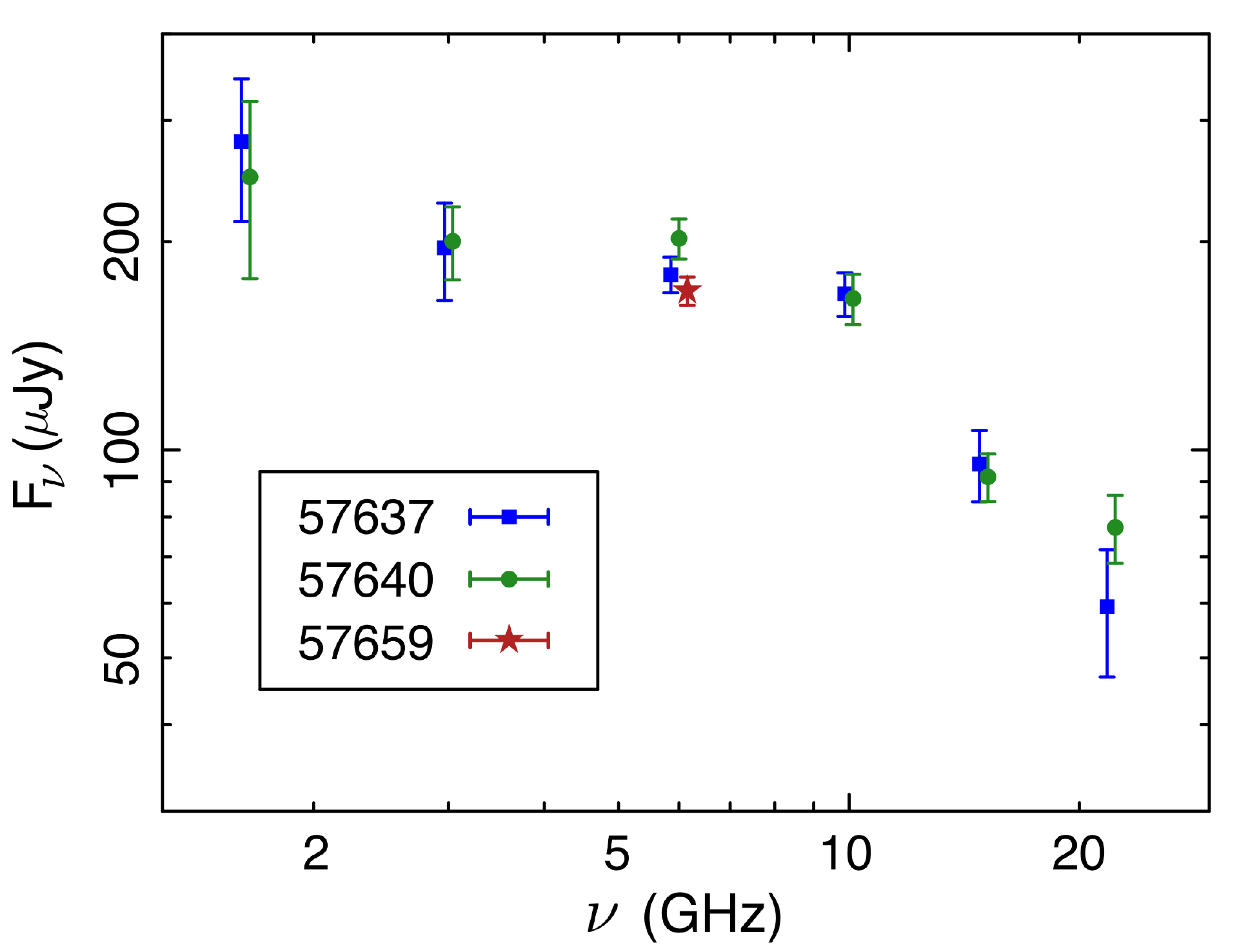}
\end{center}
\label{fig:vlaspectrum}
\end{figure}

\clearpage
\noindent
\large{\bf Supplementary Material}

\begin{addendum}

\item[Extended acknowledgements]
Aside from the Very Large Array and the Arecibo Observatory, the observational campaign reported here relied on several other telescopes and the analysis was supported by many different organizations.
S.C., R.S.W., and J.M.C. acknowledge prior support from the National Science Foundation through grants AST-1104617 and AST-1008213.
This work was partially supported by the University of California Lab Fees program under award number LF-12-237863.
The research leading to these results has received funding from the European Research Council (ERC) under the European Union's Seventh Framework Programme (FP7/2007-2013).  J.W.T.H. is an NWO Vidi Fellow and, along with C.G.B., gratefully acknowledges funding for this work from ERC Starting Grant DRAGNET under contract number 337062.
S.P.T acknowledges support from a McGill Astrophysics postdoctoral fellowship.
M.W.A. was a participant in the 2016 Research Experience for Undergraduates in Astronomy and Astrophysics at Cornell University program, supported by grant NSF/AST-1156780.
V.M.K. holds the Lorne Trottier and a Canada Research Chair and receives support from an NSERC Discovery Grant and Accelerator Supplement, from a R. Howard Webster Foundation Fellowship from the Canadian Institute for Advanced Research (CIFAR), and from the FRQNT Centre de Recherche en Astrophysique du Quebec.
B.M. acknowledges support by the Spanish Ministerio de Econom\'ia y Competitividad (MINECO/FEDER, UE) under grants AYA2013-47447-C3-1-P, AYA2016-76012-C3-1-P, and MDM-2014-0369 of ICCUB (Unidad de Excelencia `Mar\'ia de Maeztu').
L.G.S. gratefully acknowledge financial support from the ERC Starting Grant BEACON under contract number 279702 and the Max Planck Society.
Part of this research was carried out at the Jet Propulsion Laboratory, California Institute of Technology, under a contract with the National Aeronautics and Space Administration.
S.C., J.M.C., P.D., T.J.L., M.A.M., and S.M.R. are partially supported by the NANOGrav Physics Frontiers Center (NSF award 1430284).
The European VLBI Network is a joint facility of independent European, African, Asian, and North American radio astronomy institutes.
ALMA is a partnership of ESO (representing its member states), NSF (USA) and NINS (Japan), together with NRC (Canada) and NSC and ASIAA (Taiwan) and KASI (Republic of Korea), in cooperation with the Republic of Chile. The Joint ALMA Observatory is operated by ESO, AUI/NRAO and NAOJ.
Based on partial observations obtained at the Gemini Observatory, which is operated by the Association of Universities for Research in Astronomy, Inc., under a cooperative agreement with the NSF on behalf of the Gemini partnership: the National Science Foundation (United States), the National Research Council (Canada), CONICYT (Chile), Ministerio de Ciencia, Tecnolog\'{i}a e Innovaci\'{o}n Productiva (Argentina), and Minist\'{e}rio da Ci\^{e}ncia, Tecnologia e Inova\c{c}\~{a}o (Brazil).
This research has made use of the Keck Observatory Archive (KOA), which is operated by the W. M. Keck Observatory and the NASA Exoplanet Science Institute (NExScI), under contract with NASA. The data set used here for cross checking was made publicly available by PI S.~R.~Kulkarni.
Portions of the results presented were based on observations obtained with {\it XMM-Newton}, an ESA science mission with instruments and contributions directly funded by ESA Member States and NASA, and with the {\it Chandra} X-ray Observatory.  This research has made use of the NASA Astrophysics Data System (ADS) and the arXiv.

\end{addendum}

\end{document}